\documentstyle[12pt]{article}
\begin{document}
\newcommand{\beq}{\begin{equation}}
\newcommand{\eeq}{\end{equation}}
\newcommand{\bea}{\begin{eqnarray}}
\newcommand{\eea}{\end{eqnarray}}
\newcommand{\bal}{\begin{array}{ll}} \newcommand{\eal}{\end{array}}
\newcommand{\nn}{\nonumber}
\newcommand{\hl}{{\hat \lambda}}
\newcommand{\cm}{{\cal M}}
\def\Sup{\mathop{\rm\Sup}\nolimits}
\def\R {\rm R \kern -.35cm I \kern .19cm}
\def\C{ {\rm C \kern -.15cm \vrule width.5pt \kern .12cm}}
\def\Z{ {\rm Z \kern -.30cm \angle \kern .02cm}}
\def\1{ {\rm 1 \kern -.10cm I \kern .14cm}}
\def\N{ {\rm N \kern -.31cm I \kern .15cm}}

\begin{titlepage}
\begin{flushright}
hep-ph/9612220 \\December 1996 
\end{flushright}
\vskip 1cm
\centerline{\bf {A NEW INTERPRETATION OF THE PROTON-NEUTRON BOUND
STATE.}} 
\centerline{\bf {THE CALCULATION OF THE BINDING ENERGY.}} \vskip 1.5cm
\centerline{\bf{N.B. Mandache } \footnote{e.mail address : 
Bogdan.Mandache@lpgp.u-psud.fr}}
\vskip .5cm
\centerline{ Universit\'e Paris-Sud, LPGP, B\^atiment 212}
\centerline{91405 ORSAY CEDEX, FRANCE
\footnote{Permanent address :
Institute of Atomic Physics, IFTAR, Lab.22, Bucharest-Magurele,
Romania}}
\vskip 1.5cm
\begin{abstract}
We treat the old problem of the proton-neutron bound state (the
deuteron). Using a new concept of incomplete (partial) annihilation
process we derive a formula for the binding energy of the deuteron,
which does not contain any new constant. Some implications  of this new
approach are discussed.

 \end{abstract}
\end{titlepage}

\section{Introduction.}
	Since the strongest interaction between two particles is the
particle-antiparticle annihilation process, we must take into account
this process when we treat the problem of the elementary nuclear bound
state. In an annihilation process a maximum quantity of energy is
released and we know that the strengh of a bound state is   proportional
to the energy released to the exterior when the bound state is build up.

The concrete fact from which we start in this paper is the analysis of
the contradictory characteristics of the neutron. Indeed, the neutron
which has a zero total electric charge, seems to have very strange 
charge structure, having perhaps no parallel
example in whole of the hadron physics \cite{A}.

A new concept of incomplete (limited) annihilation process is used to
explain the formation of the proton-neutron bound state (paragraph 2).
The formula derived for the binding energy of this state does not
contain any new constant (paragraphs 3 and 4). Some comments are given
in paragraph 5.

\section {The incomplete annihilation process.} \vskip .5cm
The magnetic form factors of the neutron and proton have the same
functional dependence, given by a  scaling law determined expermentally,
which proves that the two particles have similar structures. But the
magnetic moment of the neutron ($\mu_n=-1.91\mu_N$) is of opposite sign
to that of the proton ($\mu_p=+2.79\mu_N$).

In spite of the fact that the electric charge of the neutron is zero,
the mean-square charge radius of the neutron, determined experimentally,
is different from zero ($<r^2_E(neutron)>=-0.112 fm^2)$ and again
opposite in sign to that of the proton ($<r^2_E(proton)>=+0.67 fm^2)$.

While the ratio of the magnetic moment of the neutron to that of the
proton was derived using the SU(6) symmetry, the calculated value (-2/3)
being in a good agreement with the experimental one, the value and the
sign of $<r^2_E(neutron)>$ are difficult to explain \cite{A}. Further
more in \cite{B} was derived a theorem proving a contradiction between
the experimental and theoretical data on the $<r^2_E(neutron)>$; more
elaborate theoretical treatments do not elucidate complete the probleme
\cite{C}.

The data on the neutron magnetic moment and $<r^2_E(neutron)>$, in
comparison with that of the proton, suggest that the neutron consists of
two hadronic masses: a larger one $m_{h^-}$ which carries a negative
electric charge and dominates into the neutron structure, and, to keep
the charge neutrality of the neutron, a smaller one $m_{h^+}$ which
carries a positive electric charge. We postulate that these two
constituent hadronic masses of the neutron obey properties similar to
the asymptotic freedom and infrared confinement.

The fact that the neutron decay is intermediate by the vectorial boson
$W^-$ suggests also that into the neutron stucture the negatively
charged mass is dominant.

The sum of these two constituent masses must be equal to the neutron mass:
\begin{equation}
m_{h^-}c^2+m_{h^+}c^2=m_{on}c^2 \label{eq:1}
\end{equation}

In the case of a perfect particle-antiparticle symmetry, for instance
proton-antiproton pair, the annihilation process of the particle with
the antiparticle is "complete" and a maximum quantity of energy is
released. It is the well known particle-antiparticle annihilation
process.

In the case of an imperfect symmetry the annihilation proces is
incomplete and the released energy is correspondingly smaller.

Otherwise spoken, in the case of complete annihilation the two partners,
which have opposite electrical charges and equal  masses, have a
maximum interaction compatibility. In the case of incomplete
annihilation, since one of the partners has a mass smaller than its
partner, the interaction compatibility decreases.

Based on these we propose a new interpretation of the formation of the
proton-neutron bound state: an incomplete (limited) annihilation process
takes place between the proton and the negatively charged mass from
neutron $(m_{h^-})$ which is smaller than the mass of the proton (see
the next paragraph). The binding energy of the proton-neutron bound
state (the deuteron) is equal with the energy released in this
incomplete annihilation process.

In the case of a symmetry  with a higher degree of "imperfection", which
means that the two partners have very different masses, the annihilation
process will be weak, the released energy being much smaller. The
process of formation of the proton-electron bound state (the hydrogen
atom) can be compared with a weak annihilation process, the proton and
the electron having opposite electric charges but very different masses
(a very low degree of symmetry, the electron mass
being of leptonic origin only).

\section {The calculation of the negatively charged mass $m_{h^-}$
from neutron.}  \vskip .5cm
We start from the relativistic expression of the energy E for a particle
of rest mass $m_o$ : 
\begin{equation}
E^2 \equiv m^2c^4=m^2_oc^4+p^2c^2 \label{eq:2}
\end{equation}
where m is the "dynamical" mass.

It is well known from quantum mechanics that when such a particle
evoluate with negative kinetic energy, this means imaginary impulse, the
energy equation becomes \cite{D} :
\begin{equation}
E^2 \equiv m^2c^4=m^2_oc^4-p^2c^2 \label{eq:3}
\end{equation}
where p  was replaced in (\ref{eq:2}) by ip; m is now smaller than the
rest mass.

Let's write similar equations for the two constituents of the neutron
$m_{h^-}$ and $m_{h^+}$. Like rest masses are used the mass of the
lightest baryon (the proton) for the negatively charged mass which
dominates into the neutron, and the mass of the lightest meson (the
pion) for the positively charged mass  from neutron:
\begin{equation}
E^2_{h^-} \equiv m^2_{h^-}c^4=m^2_{op^-}c^4-p^2_{h^-}c^2 \label{eq:4}
\end{equation}
\begin{equation}
E^2_{h^+} \equiv m^2_{h^+}c^4=m^2_{o\pi^+}c^4-p^2_{h^+}c^2 \label{eq:5}
\end{equation}

We postulate that:
\begin{equation}
p_{h^-}=p_{h^+}  \label{eq:6}
\end{equation}
taking into account the universality of the impulse conservation law.

Now is straightforward to calculate $m_{h^-}$ and $m_{h^+}$. From
(\ref{eq:1}), (\ref{eq:4}), (\ref{eq:5}) and (\ref{eq:6})  it results:
\begin{equation}
m_{h^-}c^2={m^2_{oh^-}c^2-m^2_{o\pi^+}c^2+m^2_{on}c^2\over 2m_{on}}
\label{eq:7}
\end{equation}
and an analogously relation for $m_{h^+}c^2$.

Replacing the rest mass values:
\begin{eqnarray}
m_{op^-}c^2&=&938.259 MeV  \nonumber \\
m_{o\pi^+}c^2&=&139.568 MeV  \nonumber \\
m_{on}c^2&=&939.552 MeV \label{eq:8}
\end{eqnarray}
we obtain the values of the negatively charged and positively
charged masses  from neutron:
\begin{eqnarray}
m_{h^-}c^2&=&927.893 MeV , \nonumber \\
m_{h^+}c^2&=&11.659 MeV . \label{eq:9}
\end{eqnarray}

If like in \cite{D} we take ${E_{h^+} \equiv m_{h^+}c^2 \simeq 0}$, it
results from (\ref{eq:5}) that ${p_{h^+}=m_{o\pi^+}c}$ and we obtain
directly from (\ref{eq:4}) the value of $m_{h^-}$ :
\begin{equation}
m_{h^-}c^2=\sqrt{m^2_{op^-}c^4-m^2_{o\pi^+}c^4}=927.820 MeV
\label{eq:10}
\end{equation}
The relation (\ref{eq:1}) is in this case satisfied by the contribution
of other degrees of freedom than $m_{h^+}c^2$.

The smalness of the positively charged mass from neutron (less than 12
MeV) means that it is "empty" of hadronic mass (contains only "current"
mass) and for this reason does not participate at the strong
interaction, in contrast with the negatively charged mass from neutron.

\section {The calculation of the binding energy of the deuteron.}
\vskip .5cm
Now, we  will treat quantitatively the incomplete annihilation process.
The system proton-antiproton $({m_{op^+}\over m_{op^-}}=1)$ has a
maximum interaction compatibility; the distance of "approach" between
the particle and the antiparticle at "complete" annihilation is
$a_{ann}$, where the annihilation probability is maximum \cite{E}.

The system proton-negatively charged mass from neutron $({m_{op^+}\over
m_{h^-}}>1)$ has a smaller interaction compatibility, it annihilates
incompletely, the distance of approach being x (larger).

For a mass ratio much higher than 1 $({m_{op^+}\over m_-}>>1)$, where
$m_-$ is the mass of the particle with negative charge, the system has a
much smaller interaction compatibility, the annihilation being weak.
Since in this case $m_-$ is  "empty" of hadronic mass we postulate that
the distance $(a_{m_-})$ and the binding energy $(E_{m_-})$ are given by
the known formulae from quantum mechanics for the Coulomb potential
(Bohr formulae).

Taking as parameter-the mass ratio, we systematize in the following
table the above presented dependences:

${m_{op^+}\over m_{op^-}}=1$....... $a_{ann}$.......$-{1\over
{2m_{op^+}c^2}}$

${m_{op^+}\over m_{h^-}}>1$.........$x$ ............${1\over {E_x}}$

${m_{op^+}\over m_-}>>1$......${a_{m_-}}$........${1\over {E_{m_-}}}$

In the same table are presented also the energies released in the
complete, incomplete and weak annihilation process respectively, taken
with the sign minus, this means just the binding energies of the states
formed by these processes. Since the interaction compatibility, and by
this also the binding energy, increases when the characteristic distance
of approach (interaction) decreases, in analogy also with the Bohr
formulae, we have taken the binding energies invers proportional to the
distances. For distance $a_{ann}$ (complete annihilation), the value of
the released energy is $2m_{op^+}c^2$, and this was taken as the
"binding energy".

Assuming linear dependences, from the above table (if you represent
graphically the distances in function of  the parameter-mass ratio-from
like triangles) the following relation can be drawn for distance x:
\begin{equation}
{{m_{op^+}\over m_-}-1\over {m_{op^+}\over m_{h^-}}-1}={a_{m_-}-a_{ann}
\over x-a_{ann}}  \label{eq:11}
\end{equation}
where $a_{m_-}={\hbar^2\over m_-e^2}$.

Since ${m_{op^+}\over m_-}>>1$ and $a_{m_-}<<a_{ann}$ ($ \sim 1fm$,
see\cite{E}), we obtain:
\begin{equation}
x=K_1({m_{op^+}\over m_{h^-}}-1)+a_{ann} \label{eq:12}
\end{equation}
where $K_1={\hbar^2\over m_{op^+}e^2}$.    

Likewise, from the same table we obtain for the binding energy $E_x$:
\begin{equation}
{{m_{op^+}\over m_-}-1\over {m_{op^+}\over m_{h^-}}-1}={{{1\over
E_{m_-}}+{1\over 2m_{op^+}c^2}}\over {{{1\over E_x}+{1\over
{2m_{op^+}c^2}}}}} \label{eq:13}
\end{equation}
where ${E_{m_-}=-{m_-{e^4\over 2\hbar^2}}}$.

Since ${m_{op^+}\over m_-}>>1$, we obtain:
\begin{equation}
E_x=-{1\over {({m_{op^+}\over m_{h^-}}-1)K_2+{1\over 2m_{op^+}c^2}}} 
\label{eq:14}
\end{equation}
where $K_2={2\hbar^2\over m_{op^+}e^4}$.

Substituting in (\ref{eq:14}) the values of the constants  and of
$m_{h^-}$, calculated in the paragraph 3 (the relation 9), we obtain for
the binding energy $E_x$ the value:
\begin{equation}
E_x=-2.233 MeV \label{eq:15}
\end{equation}
which, in the proposed model, is just the binding energy of the deuteron
and is in fairly good  agreement with the experimental value:
\begin{equation}
E_{exp}=-2.224 MeV \label{eq:16}
\end{equation}
Substituting in (12) the values of the constants and of $m_{h^-}$ from
(9),  we obtain:
\begin{equation}
x={0.321 \times {10^{-13}}cm+a_{ann}} \simeq {1.3fm} \label{eq:17}
\end{equation}

It is interesting to note that for  the other value of $m_{h^-}$
(relation 10) we obtain for the binding energy of the deuteron the
value;
\begin{equation}
E'_x=-2.188MeV \label{eq:18}
\end{equation}
which means that the experimental value (16) is placed between the two
calculated values (relations 15 and 18).

It should be noted that (14), if the term $1\over {2m_{op^+}c^2}$ is
neglected, which in the present case is a very good approximation, is
identical with a Bohr formula, with the fundamental distinction that the
reduced mass is of the form:
\begin{equation}
\mu={m_{op^+}m_{h^-}\over {m_{op^+}-m_{h^-}}} \label{eq:19}
\end{equation}
Indeed, the exppression of the binding energy (14) gets:
\begin{equation}
E_x=-{{m_{op^+}m_{h^-}\over {m_{op^+}-m_{h^-}}}{e^4\over 2\hbar^2}}
\label{eq:20}
\end{equation}
The value of the mass  term (19), 83.986GeV or 83.392GeV depending of
the value of $m_{h^-}$ we used (relations 9 or 10), is very near the
mass value of the charged intermediate vector W-boson (approx. 80 GeV).
Then the expression of the binding energy of the deuteron can be
written, in a precision  of 5 percents, in the surprisingly form:
\begin{equation}
E_x \simeq {-{M_W}{e^4\over {2\hbar^2}}} \label{eq:21}
\end{equation}
We could call this system formed by a negative charged vector W-boson
which evoluate in a positive, central coulombian field a "heavy atom".

 \section{Discussions.}
It is important to note that in \cite{E}  it was proved that at the
distance where the attraction is strong in the NN potential ($ \sim
1fm$), the $N\bar N$ potential is dominated by the annihilation.

It is also well known that the p-n triplet potential is different from
the p-n singlet, p-p and n-n potentials which are characterized by  an
important hard-core repulsion \cite{F}. On the contrary the p-n triplet
potential, which represent the  deuteron bound state, has a negligible
hard-core repulsion.

On the other hand it is known that the $N\bar N$ potential is
characterized by the complete lack of hard-core repulsion, the core
being strongly attractive \cite{G}. This means that the triplet p-n
bound state, described here by  an incomplete annihilation process, has an
intermediate position between the unbound nucleon-nucleon states (p-p,
n-n and p-n singlet) and the $N\bar N$ states, characterized by a
"complete" annihilation process. 

We stress that the formula derived for the deuteron binding energy,
either (14) or (21), does not contain any new constant.

In particular the formula (21) is a relation between the strong
interaction constant ${g^2_N}\over {{\hbar}c}$, characteristic to the
deuteron bound state, the nucleon mass and the constants of the
electroweak interaction $({M_W}, {e^2\over {{\hbar}c}})$. Indeed from:
\begin{equation}
E_x \simeq {-{M_W}{e^4\over {2\hbar^2}}} \equiv {-\mu_D{g^4_N\over {2\hbar^2}}} \label{eq:22}
\end{equation}
where $\mu_D$ is the deuteron reduced mass $(\mu_D \simeq {m_{op}\over
2})$ it results:
\begin{equation}
{{g^2_N}\over {{\hbar}c}}={{\sqrt{M_W\over {\mu_D}}}{e^2\over
{{\hbar}c}}} \label{eq:23}
\end{equation}

Another observation regarding the relation (21): the vectorial W-boson
has the spin 1, like the triplet  nuclear bound state.

It is to be underlined that the present approach of the proton-neutron
bound state (in particular the
derived formula 14) is based on the proportionality between the
"dynamical" mass and the strong interaction, racording the "very" strong
interaction (complete annihilation, ${m_{op^+}\over m_{op^-}}=1$) to the
e.m. interaction (atomic bound state, ${m_{op^+}\over m_-}>>1$).

\vskip 1.2cm
{\bf Acknowledgements}
\vskip .8cm
I wish to thank E.Dudas for very fruitful discussions and suggestions.

\end{document}